\documentclass{PoS}
\def\eslt{E_T^{\rm miss}}

\def\to{\rightarrow}

\def\bi{\begin{itemize}}
 \def\ei{\end{itemize}}

\def\c1p{C1^\prime}
\def\ta{\tilde a}
\def\tG{\tilde G}

\def\ta{\tilde a}

\def\tg{\tilde g}

\def\tq{\tilde q}

\def\tz{\widetilde Z}
\def\alt{\stackrel{<}{\sim}}
\def\agt{\stackrel{>}{\sim}}
\def\be{\begin{equation}}  
\def\ee{\end{equation}}  
\def\bea{\begin{eqnarray}}  
\def\eea{\end{eqnarray}}  

\newcommand\annp[3]{{\it Annals\ Phys.\ }{\bf #1} (#2) #3}
\newcommand\sjp[3]{{\it Sov.\ J.\ Nucl.\ }{\bf #1} (#2) #3}
\newcommand\prd[3]{{\it Phys.\ Rev.\ }{\bf D #1} (#2) #3}

\newcommand\prl[3]{{\it Phys.\ Rev.\ Lett.\ }{\bf #1} (#2) #3}
\newcommand\plb[3]{{\it Phys.\ Lett.\ }{\bf B #1} (#2) #3}
\newcommand\jhep[3]{{\it J. High Energy Phys.\ }{\bf #1} (#2) #3}
\newcommand\jcap[3]{{\it JCAP\ }{\bf #1} (#2) #3}

\newcommand\npb[3]{{\it Nucl.\ Phys.\ }{\bf B #1} (#2) #3}
\newcommand\epjc[3]{{\it Eur.\ Phys.\ J. }{\bf C #1} (#2) #3}
\newcommand\ptp[3]{{\it Prog.\ Theor.\ Phys.\ }{\bf #1} (#2) #3}
\newcommand\zpc[3]{{\it Z.\ Physik }{\bf C #1} (#2) #3}

\newcommand\njp[3]{{\it New\ Jou.\ Phys.}{\bf  #1} (#2) #3}

\title{Theoretical expectations for dark matter detection\\
 at the LHC}

\ShortTitle{Dark matter at LHC}

\author{\speaker{Howard BAER}\\
        Dep't of Physics and Astronomy\\
        University of Oklahoma\\
        Norman, OK, 73019, USA\\
        E-mail: \email{baer@nhn.ou.edu}}

\abstract{I present an overview of theoretical expectations for
detection of dark matter (DM) at the LHC, concentrating entirely on 
supersymmetric candidates. En route, I present a unified theory 
which explains dark matter signals at indirect and direct detection
experiments. 
While direct DM detection is likely impossible at LHC, detection
of SUSY matter states is robust, where LHC with $\sqrt{s}=7$ TeV and 
1fb$^{-1}$ can access $m_{\tg}\sim 1$ TeV for $m_{\tq}\sim m_{\tg}$,
and $m_{\tg}\sim 640$ GeV for $m_{\tq}\gg m_{\tg}$.
Models with well-tempered neutralinos will soon be tested by Xenon-100/LUX,
and should provide a distinctive mass edge at LHC in the $m(\ell^+\ell^- )$
distribution. 
In the case of SUSY, neutralino dark matter now seems highly disfavored
by both the magnitude of the dark matter density, and also the gravitino 
problem. Alternatively, the PQMSSM yields a solution to the strong CP problem,
provides all the benefits of SUSY, and solves several cosmological
problems. Predictions for SUSY particles at the LHC 
from the PQMSSM are quite different from the case of neutralino 
dark matter. Compelling models such as Yukawa-unifed SUSY or
Effective SUSY, which would be excluded with neutralino DM, 
are perfectly viable with mixed axion/axino DM.}

\FullConference{Identification of Dark Matter 2010\\
		July 26-30, 2010\\
		University of Montpellier 2, Montpellier France}

\begin{document}

\section{Introduction}

The astrophysical evidence for the existence of dark matter
in the universe is now overwhelming, and comes from numerous disparate
sources. Already, we know many of the properties of the putative
dark matter particle(s): it must be massive, electrically neutral, 
and predominantly cold (non-relativistic). Of all the particle states
in the Standard Model (SM), only neutrinos seem to have the first
two of these properties. However, neutrinos constitute {\it hot}
dark matter, so some other matter state is needed: the existence of
dark matter requires new physics beyond that of the Standard Model.

In the theory literature, there exist numerous candidate states
that might make up the dark matter: Kaluza-Klein (KK) photons or KK
gravitons, lightest $T$-parity odd particles from Little Higgs 
theories, branons, Q-balls etc. However, dark matter emerges
naturally from two quite different theories which solve long standing
problems in particle physics. 

The first of these, supersymmetry or SUSY\cite{wss}, has been invoked to 
stabilize the hierarchy problem wherein scalar masses such as 
for the Higgs boson tends to blow up to the largest scale 
present in the theory.
When the SM is supersymmetrized, the quadratic divergences are tamed; The resulting
softly broken theory, the MSSM, provides a bonafide WIMP candidate-- the lightest
neutralino $\tz_1$. If local SUSY, or supergravity (SUGRA) is invoked, 
then the gravitino may also play the role of dark matter.

The other compelling theory is the Peccei-Quinn-Weinberg-Wilczek (PQWW)
solution to the strong $CP$ problem\cite{pqww}. t'Hooft's solution to the
QCD $U(1)_A$ problem suggests that the QCD Lagrangian should contain
a $CP$ violating $\frac{\bar{\theta}}{32\pi^2}F\tilde{F}$ term, which 
leads to large contributions to the neutron EDM. 
But experiment tells us that $\bar{\theta}<10^{-11}$\cite{axreview}. PQ suggested an
additional broken $U(1)_{PQ}$ global symmetry. The resulting Goldstone 
boson-- the axion-- provides additional field dependent $F\tilde{F}$ terms
in the QCD Lagrangian. The axion field relaxes to the minimum of its
potential, causing the entire $F\tilde{F}$ term to dynamically go to zero.
In the process, coherent oscillations of the axion field fill the universe
with non-relativistic axions, which are excellent candidates for CDM\cite{absik}.

Of course, these two compelling theories-- SUSY and PQWW-- 
are not mutually exclusive, and actually enhance each other: one can
build a PQMSSM theory\cite{pqmssm} which may contain axions $a$ as well as 
$R$-parity odd axinos $\ta$. Thus, dark matter in the PQMSSM is not just
one particle; one may have mixed axion-axino CDM, mixed axion-gravitino
CDM or mixed axion-neutralino CDM. 

As we will see, there are dark clouds on the horizon for
theories with pure neutralino or pure gravitino CDM. However, 
mixed axion-ino CDM seems to work just fine! 
This has rather large implications for what LHC and dark matter detection
experiments may find.

\section{SUSY at the LHC}

Due to time limits, I will focus here on gravity-mediated SUSY breaking
models (SUGRA), 
wherein the gravitino is expected to be around the weak scale, and
to set the mass scale for the other SUSY particles. Gauge mediation 
(GMSB) doesn't seem to naturally yield CDM. 
Anomaly mediation (AMSB) can yield the correct CDM abundance: 
for a recent analysis, see \cite{shibi}.

The paradigm model for SUSY phenomenology is the minimal supergravity
or mSUGRA model\cite{an} (also called the CMSSM by some authors). The mSUGRA
model features the well-known parameter space 
\be
m_0,\ m_{1/2},\ A_0,\ \tan\beta ,\ sign(\mu ) .
\ee
One may stipulate the GUT scale parameters $m_0$, $m_{1/2}$, $A_0$ 
and $B_0$, and run them via RGEs down to the weak scale, and calculate all
sparticle masses and mixings. 
The value of $B$ is traded for $\tan\beta$ upon scalar potential minimization.
We adopt the Isajet subprogram
Isasugra\cite{isasugra} for this purpose. Once the physical masses
and couplings are known, decay widths and production cross sections 
may be calculated. 
These are all encoded in Isajet\cite{isajet}.
The neutralino relic density is calculated with IsaReD\cite{isared}.

Assuming neutralino dark matter, one may calculate $pp\to \tz_1\tz_1$
production. However, there is nothing in the final state for detectors
to trigger on, unless a hard QCD jet is radiated from the initial state.
Thus, direct dark matter production is quite useless at LHC.
Instead, for SUSY theories, one wants to produce the {\it other}
matter states associated with the new physics, which may decay later into
DM states.
For LHC-- a hadron collider-- the strongly interacting states--
the squarks and gluinos-- usually have the largest cross sections.
These are frequently also the heaviest states, so they are expected 
to decay via a {\it cascade}\cite{cascade} into the lightest SUSY
particle (LSP) plus a variety of hard jets and hard (isolated) leptons\cite{multi}.

LHC has turned on at $\sqrt{s}=7$ TeV in 2010, and will continue running 
at this energy in 2011. Total SUSY production cross sections
at $\sqrt{s}=7$ TeV are shown in Ref. \cite{lhc7}. In the case
where $m_{\tq}\sim m_{\tg}$, these cross sections range to
well over $10^4$ fb for $m_{\tg}\sim 400-500$ GeV. As of December, 2010,
Atlas and CMS each have $\sim 0.045$ fb$^{-1}$ of integrated luminosity, 
so there already could be hundreds of SUSY events lurking in their data!
 
The $\tg$ and $\tq$ decay usually through a cascade involving several
steps into jets, leptons plus $\eslt$. Combining distinct sparticle
production cross sections with the many decay possibilities yields
of order $10^5$ distinct $2\to n$ subproceses. The way to make sense 
out of these is to embed them into an event generator program, so
the expected SUSY events can be generated probabilistically, and 
including QCD radiation, hadronization, and treatment of 
the underlying event. 

We break the expected signatures up according to the presence of
isolated leptons: jets$+\eslt$, $1\ell$ +jets$+\eslt$, 
opposite sign dileptons (OS) plus jets$+\eslt$, same-sign dileptons
plus jets $+\eslt$, $3\ell$ plus jets $+\eslt$, etc. 
Standard Model processes (backgrounds, BG) which yield the same signatures
must also be computed. Judicious cuts must be made to select signal
events from BG. We invoke a multi-dimensional grid of cuts so as 
to optimize signal over BG in various regions of parameter space.
By requiring signal $S$, for a given integrated luminosity (IL), 
to exceed $max[5\sigma , 5\ events,\ 0.2\times BG]$, 
we can determine the reach of LHC for SUSY in the different multi-lepton channels.

\begin{figure}[tbh]
\includegraphics[width=9cm]{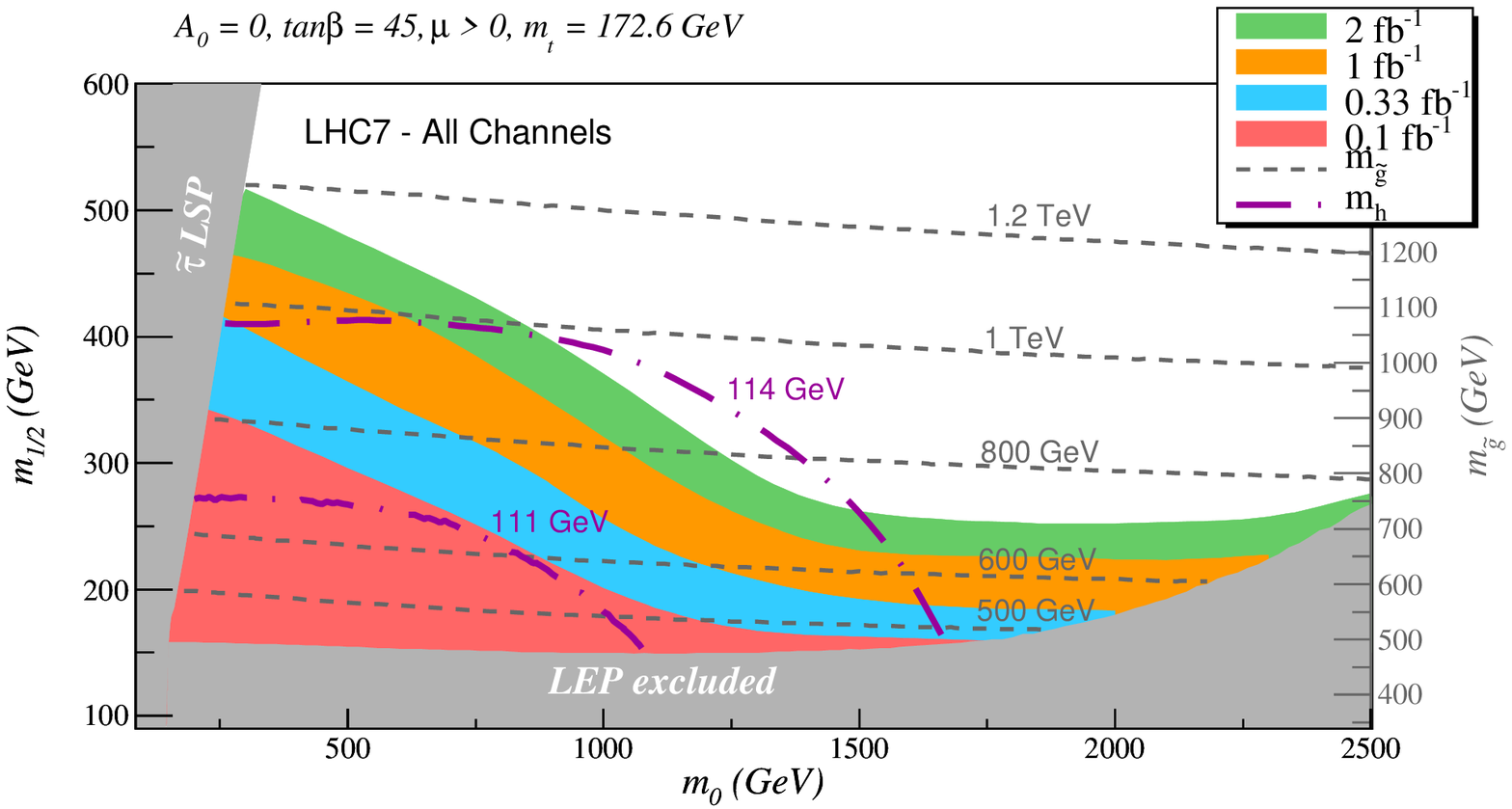}\\
\includegraphics[width=9cm]{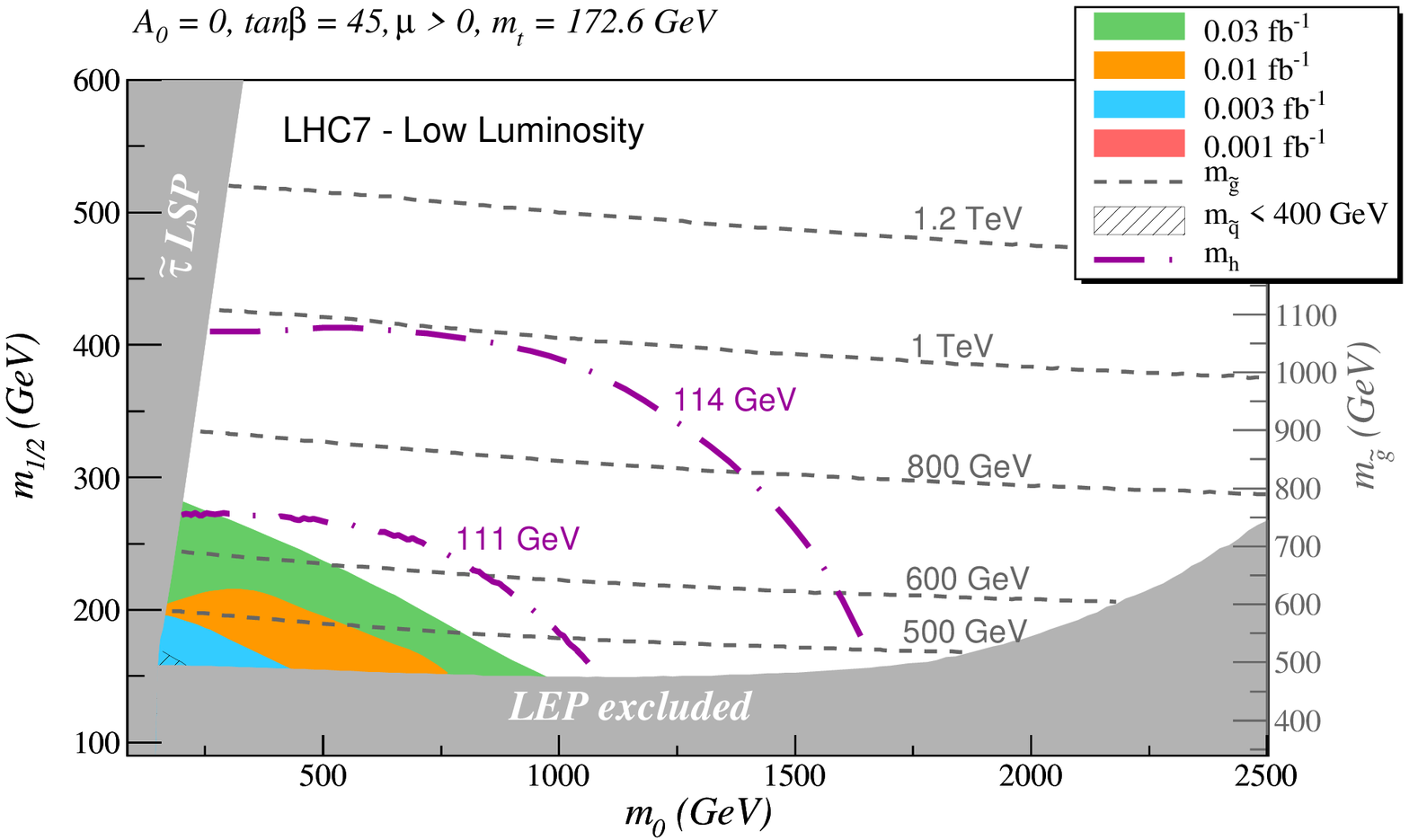}
\caption{Upper frame: Reach of LHC7 for SUSY in the mSUGRA model
with $\sqrt{s}=7$ TeV and integrated luminosity ranging from
100-2000 pb$^{-1}$.
Lower frame: Reach of LHC7 for lower values of integrated luminosity.
}\label{fig:lhc7}
\end{figure}

The results from Ref. \cite{lhc7} are shown in Fig. \ref{fig:lhc7}.
Even for $IL=0.1$ fb$^{-1}$, the LHC7 reach extends far past 
Tevatron bounds: to $m_{\tg}\sim 800$ GeV for $m_{\tq}\simeq m_{\tg}$!
For 1 fb$^{-1}$, then LHC reach is to over $m_{\tg}=1$ TeV for
$m_{\tg}=m_{\tq}$, and to $m_{\tg}\sim 630$ GeV for $m_{\tq}\gg m_{\tg}$.

We will learn a LOT in 2011 about where SUSY is, or is not! 

\section{LHC versus direct/indirect detection of neutralinos}

If neutralinos are the dark matter, then they may be detected
via direct DM detection (DD) experiments such as Xenon-100 or
LUX, or indirectly (IDD) via neutrino telescopes 
(searching for neutrinos from WIMP annihilation in the solar core), 
or by space-based antimatter (Pamela or AMS) 
or gamma ray detectors (Fermi-LAT).

\begin{figure}[tbh]
\begin{center}
\includegraphics[width=7cm]{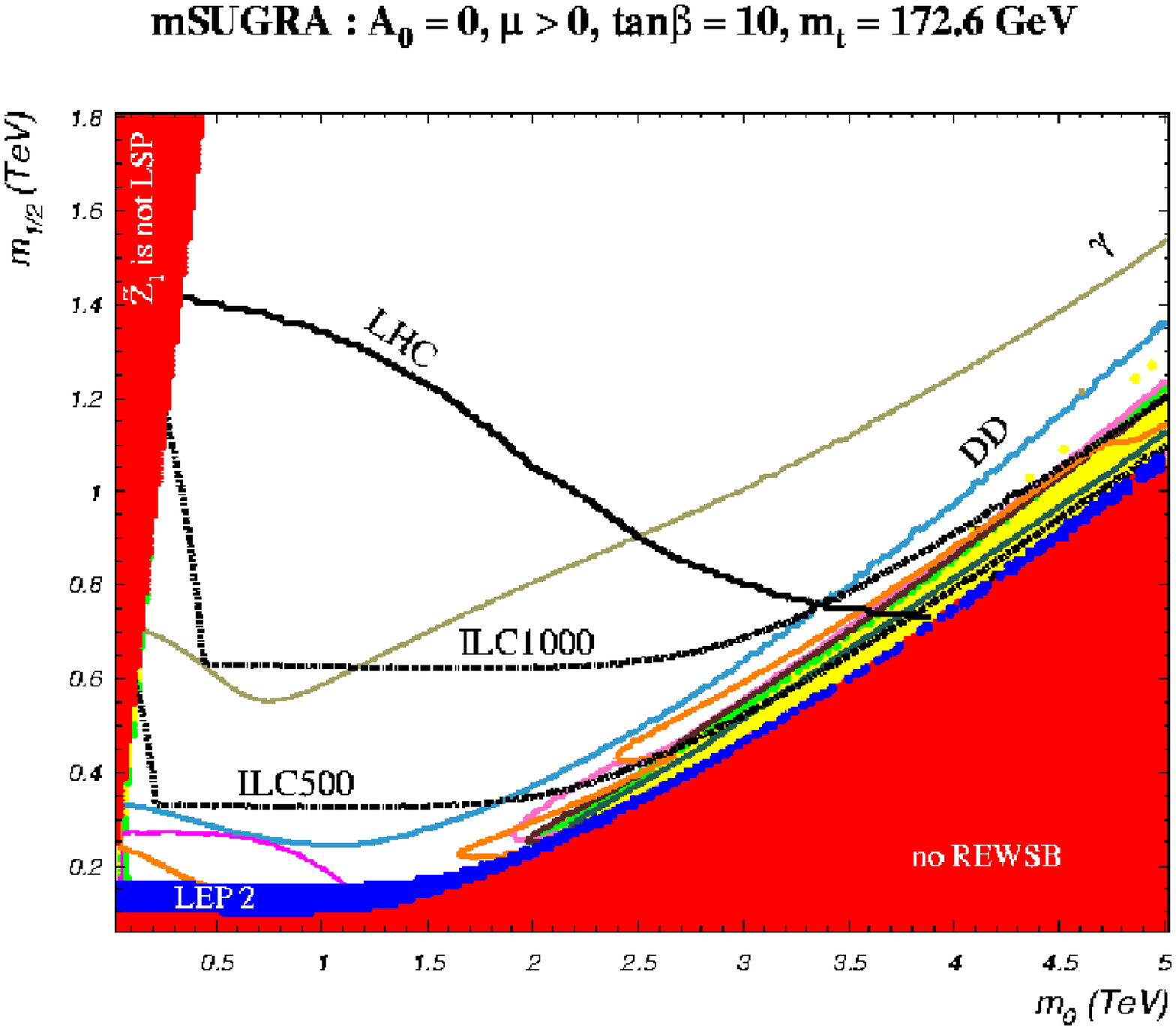}
\includegraphics[width=7cm]{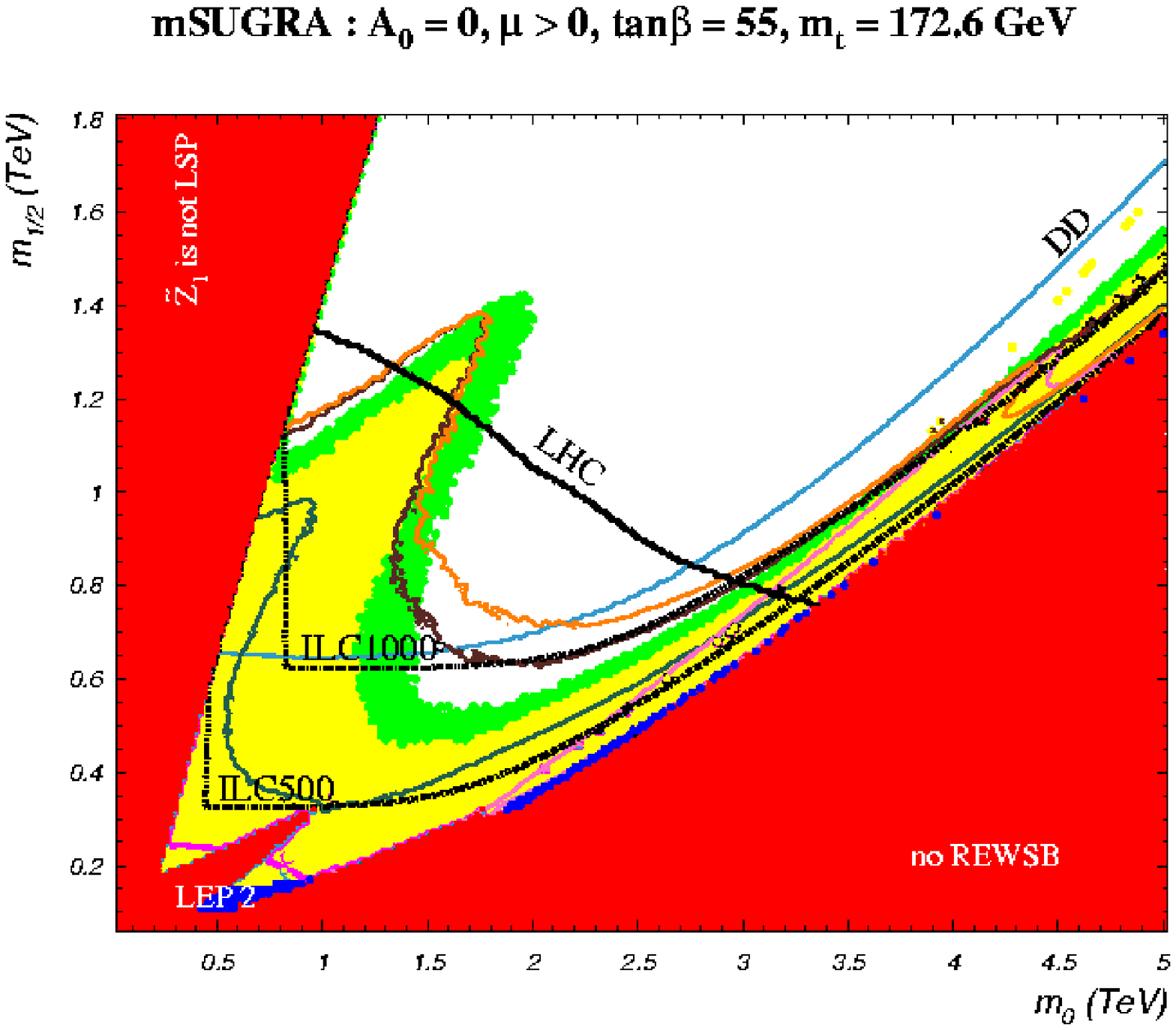}
\end{center}
\vspace{-10mm}
\begin{center}
\includegraphics[width=9cm]{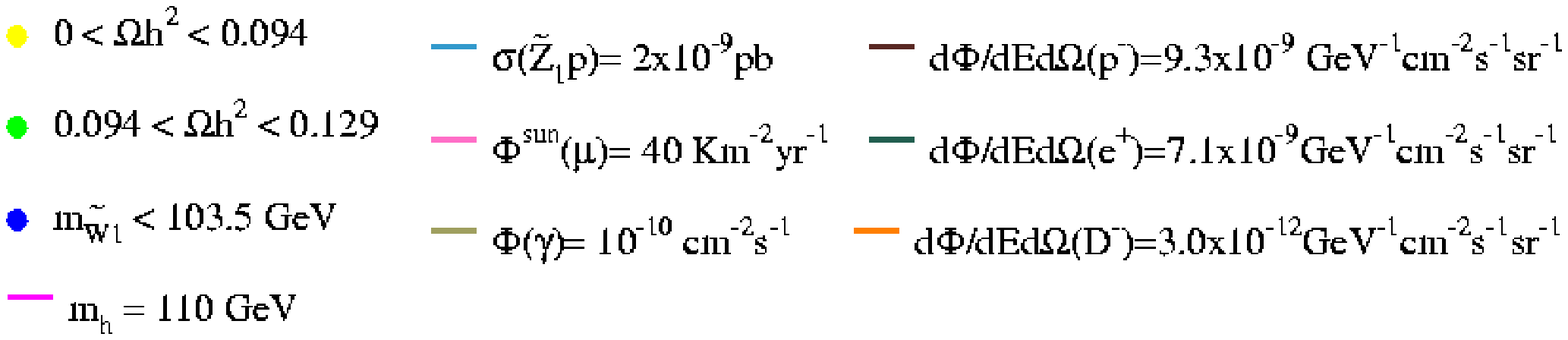}
\end{center}
\vspace{-3mm}
\caption{\small\it The projected reach of various colliders, direct and
  indirect dark matter search experiments in the $m_0\ vs.\ m_{1/2}$
  plane of the mSUGRA model for $A_0=0$, $\mu >0$, $m_t=172.6$ GeV for
  $\tan\beta =10$ (left frame) and $\tan\beta =55$ (right frame). 
  For the ID results, we   have adopted the N03 DM halo density profile.}
\label{fig:compare} 
\end{figure}

We have calculated various DD and IDD signal rates in the mSUGRA model
in Ref's \cite{bbko} and \cite{njp}, and compared to the reach of LHC14 and ILC500 and ILC1000. 
The results are shown in Fig. \ref{fig:compare}. 
Here, we see DD is largest when
$m_0$ and $m_{1/2}$ are small (low squark masses enhance squark exchange) 
or in the hyperbolic branch/focus point region where the neutralino
exists as mixed bino-higgsino state and has a large Higgs exchange
cross section. Since squarks are extremely heavy in the HB/FP region, 
this region has only limited coverage by LHC (up to $m_{\tg}\sim 1.4$
TeV for 100 fb$^{-1}$, while Xenon-100/LUX will likely cover the {\it entire} 
HB/FP region!
For IDD, neutralino annihilation cross sections are also
enhanced in the HB/FP region, as well as in the $A$-resonance region.

\subsection{Unified explanation of current DD/IDD signals}

While I am on this topic, I note that a lot of excitement has been
generated over the Pamela/ATIC/Fermi IDD signals, and also the
DAMA/LIBRA, Cogent and CDMS DD signal events. I propose a unified
explanation of these phenomena, by introducing a doublet of strongly
interacting particles $(p,\ n)$, where $p$ has charge $+1$ and $n$ is
neutral. Their putative mass is around the GeV scale. The $p$s
can be produced in various cosmic processes, and propagate a long ways:
their flux should be much greater than that of positrons, and the 
gap between the two increases with energy. Since they have the same
charge as $e^+$, they will occasionally be mistaken as a positron--
and more often at higher energies.

In the case of $n$ particles, as emphasized by Ralston\cite{ralston}, 
these can be produced by muon-induced 
radioactivity even at great depths, with some seasonal dependence 
(due perhaps to density changes in the atmosphere in summer versus
winter). Their cross sections for absorption and propagation
vary wildly, and are ill-measured in the keV regime. Unless one
invokes fiducialization by having a large mass detector with low BG, 
then some of these are expected to yield WIMP-like events.

\section{Well-tempered neutralinos (WTN)}

\begin{figure}[tbh]
\includegraphics[width=8cm]{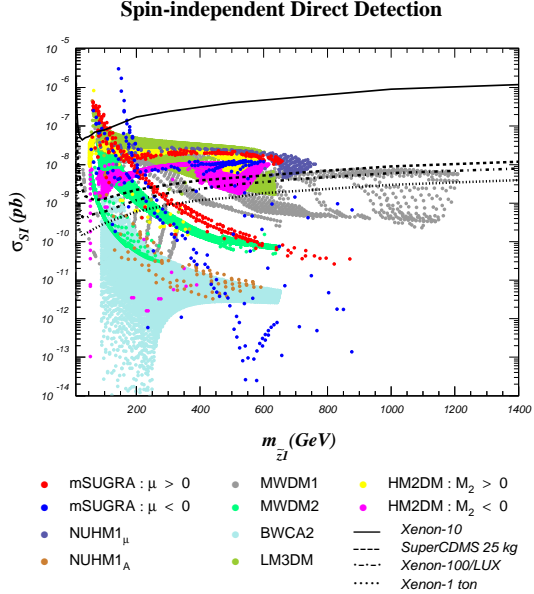}
\caption{
Predictions for $\sigma_{SI}(\tz_1 p)\ vs.\
m_{\tz_1}$, generally regarded as the figure of merit for direct
detection experiments, in various models with $A_0=0$ and $m_t=171.4$~GeV.
The special parameter of
each non-universal SUGRA model has been dialed to yield
$\Omega_{\tz_1}h^2\simeq 0.11$.  
We fix $\tan\beta=10$ except for the mSUGRA model where we allow 
$\tan\beta=10$, 30, 45, 50, 52 and 55.
We also show the projected reach of selected direct detection experiments.
}\label{fig:wtn}
\end{figure}

In Fig. \ref{fig:wtn}, we show the spin-independent $\tz_1 p$ cross
section versus $m_{\tz_1}$ for a large number of one-parameter
extensions of mSUGRA, where the GUT scale universality between matter
scalar and Higgs scalar mass parameters, or between the three gaugino
mass parameters is relaxed in a systematic way.  The details of the
various models are not essential for our present purpose, but may be
found in Ref.~\cite{wtn}.  In each such model, shown by a
different colour on the plot, this additional parameter is adjusted so
that the lightest neutralino (assumed to be the LSP) {\it saturates} the
observed relic abundance of CDM.  We also include the mSUGRA model. To
make this plot, we randomly generated points in the parameter space for
each model, and plotted it on the figure if all current collider
constraints on sparticle masses are satisfied. We also show the
sensitivity of current experiments together with projected sensitivity
of proposed searches at superCDMS, Xenon-100, LUX, WARP and at a
ton-sized noble liquid detector.  The key feature to note is that while
the various models have a branch where $\sigma_{\rm SI}(p\tz_1)$ falls
off with $m_{\tz_1}$, there is another branch where this cross-section
asymptotes to $\sim 10^{-8}$~pb\cite{wtn,pran}.  This branch
(which includes the HB/FP region of mSUGRA) includes {\it many} models
with MHDM which easily accommodate the measured relic density via {\it
tempering} of the neutralino's higgsino content\cite{arkani}.  In these cases, the
spin-independent DD amplitude -- which is mostly determined by the Higgs
boson-higgsino-gaugino coupling -- is large because the neutralino has
both gaugino and higgsino components.
The exciting thing is that the experiments currently being deployed--
such as Xenon-100, LUX, WARP and superCDMS -- will have the 
necessary sensitivity to probe this {\it entire class of models}!

A key feature of the well-tempered neutralino models is that $\mu$ is
comparable to gaugino mass $M_1$ at the weak scale so that we have a mixed 
bino-higgsino WIMP. This also implies the mass gap $m_{\tz_2}-m_{\tz_1}$ is
in the range $\sim 20-90$ GeV, which means in turn that $\tz_2\to\ell^+\ell^-\tz_1$
is a dominant decay mode, and should give rise to a visible mass edge with
shape of $m(\ell^+\ell^- )$ characteristic of a three-body decay at the LHC\cite{wtn}. 

\subsection{Smoking gun LHC signature 
for low-mass SUSY or well-tempered neutralinos}

The decay $\tz_2\to\ell^+\ell^-\tz_1$ should occur with a large branching fraction
in SUGRA models with gaugino mass unification if $m_{\tg}\alt 630$ GeV, or in WTN models.
The distribution rises to a sharp kinematic edge at $m_{\tz_2}-m_{\tz_1}$, 
which can lie between the photon and $Z$ poles\cite{mlledge}. This signature should be robust
in OS-dilepton plus jets $+\eslt$ events at LHC, provided the above conditions hold, and
that there is not accidental negative interference between slepton and $Z$ exchange
in the three body decay amplitudes.

\begin{figure}[tbh]
\includegraphics[width=7cm]{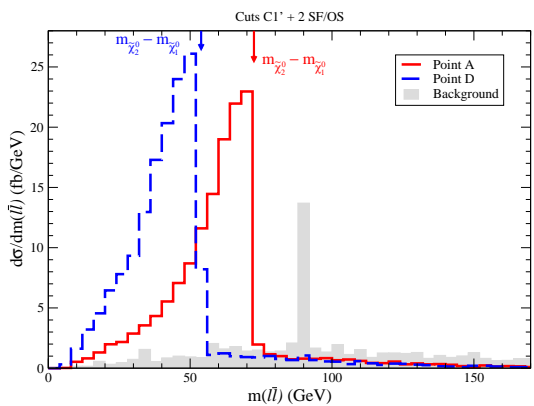}
\caption{SF/OS dilepton invariant mass distribution after cuts $C_1'$ 
from benchmark points A (full red line) and D (dashed blue line)
along with SM backgrounds, for two Yukawa-unified benchmark
points, from Ref. \cite{so10lhc}.
}\label{fig:mll}
\end{figure}

\section{Two problems for neutralino DM}

The idea of CDM consisting of neutralino WIMPs is a popular one.
Much has been made lately of the so-called ``WIMP miracle'', 
wherein it is claimed that WIMPs have exactly the right properties
to generate the observed abundance of dark matter, provided they 
undergo weak interactions and have weak scale masses. 
However, this argument applies better to a hypothetical
left-hand neutrino with mass around the weak scale (perhaps in a 
fourth generation extension of the SM). Such massive left-hand neutrinos
have long been ruled out because their direct detection cross sections
are large.

\begin{figure}[tbh]
\includegraphics[width=7cm,angle=-90]{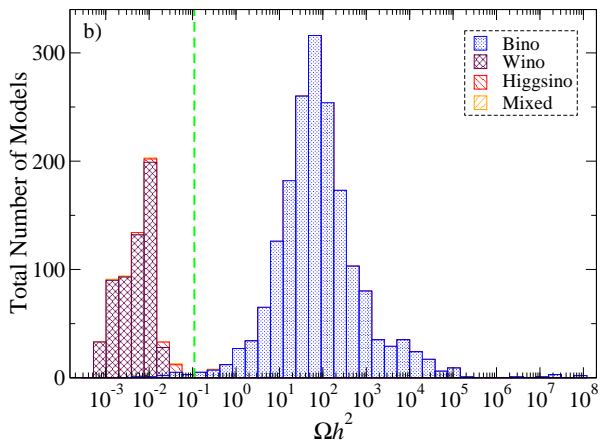}
\caption{
Projection of the number of models generated by a linear scan over
SUGRA-19 parameters, versus neutralino relic density $\Omega_{\tz_1}h^2$.
Models with mainly bino, wino, higgsino or a mixture are indicated by the 
various color and symbol choices.
We require only models with $m_{\tz_1}<500$ GeV to avoid
too large of fine-tuning in the SUSY parameters
}\label{fig:ax19}
\end{figure}

When one plots the thermal neutralino relic density in the 
$m_0\ vs.\ m_{1/2}$ plane of the mSUGRA model, it is striking that almost all 
parameter space is ruled out, save for a few narrow regions: stau-coannihilation,
$A$-resonance annihilation, HB/FP and the $h$-resonance. The relic density more typically
in mSUGRA space is of order $\Omega_{\tz_1}h^2\sim 1-100$: far too large.
Rather high fine-tuning of parameters is needed to obtain the correct thermal
abundance\cite{eo,bbox}.
One might object that this argument applies to mSUGRA, and other models with non-universal 
soft terms have more options to get the right relic density. In Ref. \cite{ax19}, we scanned over
SUGRA models with 19 free parameters, and plotted the relic density. The result is shown in
Fig. \ref{fig:ax19}, where we have made a linear scan over 19 parameters, and required
$m_{\tz_1}<500$ GeV to avoid too much electroweak fine-tuning. From the figure, it is clear
that models with a bino-like neutralino should have $\Omega_{\tz_1}h^2\sim 1-1000$, 
while models with a wino-like or higgsino-like neutralino should have $\Omega_{\tz_1}h^2\sim 0.001-0.01$:
either far too much or far too little dark matter. The measured abundance falls exactly in the most improbable
region, which requires high fine-tuning to get a well-tempered neutralino, or just the right
sparticle mass combinations to get resonance or co-annihilation.

Another issue for the neutralino DM picture is the gravitino problem\cite{gravprob}.
If gravitinos indeed set the mass scale for SUSY particles, and have weak scale masses, 
then they should be produced at rather high rates in the early universe, 
depending on the re-heat temperature $T_R$. 
The gravitinos will undergo late-time cascade decays to a neutralino LSP which 
will destroy the light element abundances built up in BBN, unless $T_R\alt 10^5$ GeV\cite{kl}. 
If we accept $T_R<10^5$ GeV, it turns out this is not enough to support thermal
leptogenesis which requires $T_R\agt 2\times 10^9$ GeV\cite{buch}; 
and thermal leptogenesis seems favored by the emergent picture of neutrino masses via
the see-saw mechanism.
By requiring $m_{\tG}\agt 5$ TeV, the gravitino decays more quickly, and the bound\cite{moroi}
on $T_R$ goes up to $\sim 10^8$ GeV; this is enough to support at least non-thermal leptogenesis, wherein
right-hand neutrinos are produced via other mechanisms, perhaps inflaton decay\cite{ntlepto}.
But then one should add into the abundance of neutralinos those that are produced from a high rate 
of gravitino production, which exacerbates the relic density problem for binos.

One might hypothesize the gravitino to be the LSP so it is stable and doesn't suffer
late decays. But then if $\tz_1$ is NLSP, it will undergo late decays to gravitino plus hadrons, and
re-introduce the gravitino problem.

Nature may be trying to tell us something here.

\section{Mixed axion/axino CDM}

Up to now, we have ignored the strong $CP$ problem. The Peccei-Quinn
solution with an ``invisible axion''\cite{ksvz,dfsz}, is still the most attractive solution after 
over 30 years. In a SUSY context (the PQMSSM)\cite{pqmssm}, the axion is but one element of the 
axion/axino superfield, which includes along with the axion, 
the spin-${1\over 2}$, $R$-parity-odd axino
which may play the role of LSP, and the spin-0 $R$-even saxion, 
which acquires a SUSY breaking mass.
The axino mass is model dependent, and can lie anywhere from the keV-TeV range.
If $m_{\ta}< 0.2$ keV, then they could be produced in thermal equilibrium, but would
constitute HDM\cite{rtw}. If their mass is $\agt$ 100 keV, then they can still be produced
thermally (TP), but would constitute CDM\cite{ckkr}. Axino LSPs can also be produced from 
neutralino decays (NTP); since each neutralino decays to one axino, these non-thermally produced
axinos inherit the neutralino number density, and reduce the neutralino
abundance by a factor $m_{\ta}/m_{\tz_1}$. For $m_{\ta}\sim$ MeV scale, this factor
is tiny, and essentially wipes out the NTP axino abundance. The remaining abundance
comes from a mixture of TP axinos and axions produced from vacuum mis-alignment.
While the axino abundance decreases with PQ breaking scale $f_a/N$ due to
decreasing coupling strength, the axion abundance increases with $f_a/N$, due to a
likely larger initial axion field strength value.

\begin{figure}[tbh]
\includegraphics[width=8cm]{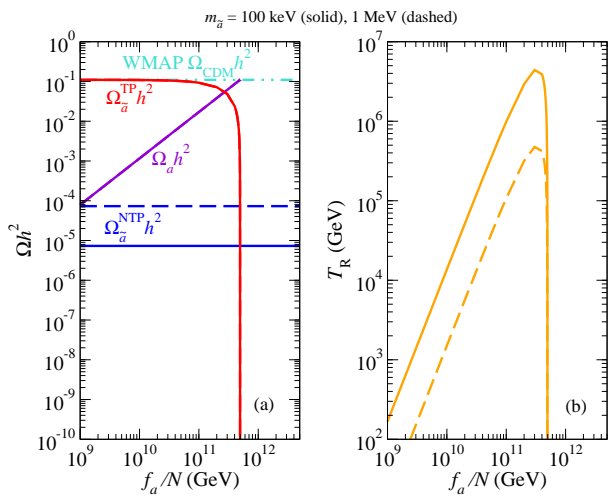}
\caption{Axion and TP and NTP axino contributions to
dark matter density for $m_{\ta}=100$ keV
versus PQ breaking scale $f_a/N$.
}\label{fig:Oh2ata}
\end{figure}

The situation is shown in Fig. \ref{fig:Oh2ata}, where we plot the component 
contributions of mixed axion/axino CDM, requiring the total abundance
$\Omega_{a\ta}h^2=0.11$. As $f_a/N$ increases, the TP axino contribution would like to decrease, 
but insisting on the measured total abundance requires an increased $T_R$ to compensate.
In this way, in frame {\it b})., we see that $T_R$ can increase into the $10^7$ GeV range, 
sufficient enough to support non-thermal leptogenesis. 
Models with a very heavy gravitino\cite{tr}, or the Asaka-Yanagida scenario\cite{AY} with
$m(sparticle)>m(gravitino)>m(axino)$ can even allow for $T_R\sim 10^{10}-10^{11}$ GeV, 
enough to support thermal leptogenesis!
The highest $T_R$ values come when $f_a/N$ is quite large, $\sim 10^{12}$ GeV, when the
mixed axion/axino dark matter is {\it mainly axion} CDM\cite{axdm}.
%

\begin{figure}[tbh]
\includegraphics[width=9cm]{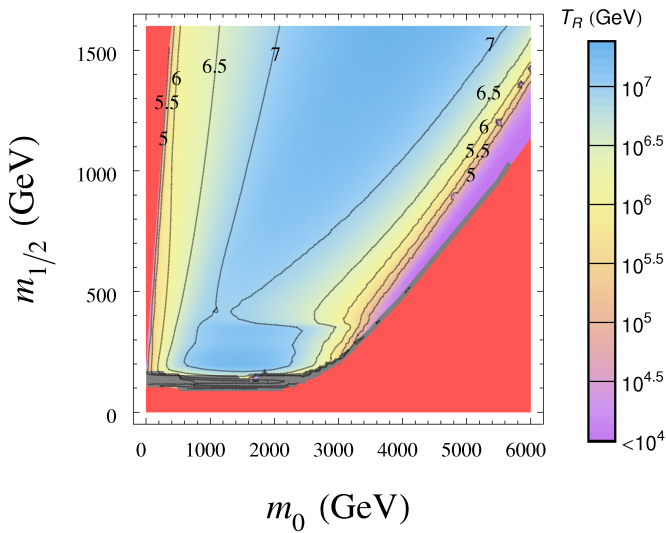}
\caption{
Contours of constant $T_R$ in the $m_0\ vs.\ m_{1/2}$
plane for $A_0=0$, $\tan\beta =10$ and $\mu >0$.
We assume $\Omega_ah^2=0.11$, and 
$\Omega_{\ta}^{TP}h^2=0.006$ and $\Omega_{\ta}^{NTP}=6\times 10^{-6}$.
}\label{fig:TRplane}
\end{figure}

For the mSUGRA model, we can adopt a rather large value of $f_a/N$ so that
we get mainly axion CDM: $\Omega_ah^2\simeq 0.11$, with 
$\Omega_{\ta}^{TP}h^2=0.006$ and $\Omega_{\ta}^{NTP}=6\times 10^{-6}$.
Then in Fig. \ref{fig:TRplane}\cite{axdm}, we plot contours of $\log_{10}T_R$. 
We see that the regions with largest $\Omega_{\tz_1}h^2$ which would be
severely excluded for $\tz_1$ DM are now perfectly viable, and can yield $T_R$ over
$10^7$ GeV: the regions of mSUGRA space which are most highly disfavored
by neutralino CDM are {\it most favored by mainly axion CDM}! A lesson: LHC searches for SUSY
should not restrict themselves to so-called neutralino dark-matter-allowed regions!
%

\section{Yukawa-unified SUSY and the LHC}

Armed with the idea of mixed axion/axino CDM, many compelling models which
gave rise to far too much neutralino CDM can now be perfectly viable.
One case is Yukawa unified SUSY, which is inspired by $SO(10)$ models with
$t-b-\tau -\nu_\tau$ Yukawa coupling unification at $M_{GUT}$. These models require $m_{16}$, 
the mass of all matter scalars, in the $\sim 10$ TeV range, while $m_{1/2}$ is very low\cite{so10}.
One obtains a pure bino-like neutralino whose annihilation rate is severely suppressed
by multi-TeV scalar exchange: $\Omega_{\tz_1}h^2\sim 10^2-10^4$. However, if we invoke
mixed axion/axino CDM, the models become perfectly viable, with cases including mainly axion
CDM as most favorable\cite{bhkss}. The prediction of Yukawa-unified SUSY is that
$m_{\tg}\alt 500$ GeV. LHC experiments will cover this range in 2011, thus ruling in or ruling out
this class of models\cite{so10lhc}.

Another example is Effective SUSY, a model by Cohen, Kaplan and Nelson\cite{ckn}, which also involves
multi-TeV scalars and too much neutralino CDM: this model is now perfectly viable
with mixed axion/axino CDM\cite{esusy}.

\section{Conclusions}

My conclusions as a bullet list:
\bi
\item The role of LHC in dark matter searches: produce matter
states associated with the dark matter, which cascade decay into 
DM states. The usual signature is multi-jet plus multi-lepton plus
$\eslt$ events, which might be accompanied by quasi-stable particle states.
\item In the case of WIMP dark matter, additional signals from
direct and indirect detection experiments should provide complementary
information.
\item Th Xe-100 experiment will soon test SUSY models with 
{\it well-tempered neutralinos}, which have DD cross sections at
$\sim 10^{-8}$ pb; these models should exhibit a $m(\ell^+\ell^- )$ mass edge
below 90 GeV at LHC.
\item Neutralinos suffer from tending to predict too big (bino-like)
or too small (wino or higgsino-like) a relic density. They also
suffer from the gravitino problem.
\item Ditto for gravitinos
\item Mixed axion/axino CDM avoids both these isues, plus
includes a solution to the strong CP problem
\item Very compelling models such as $SO(10)$-inspired 
Yukawa-unified SUSY yield far too much neutralino dark matter. However,
these models are perfectly viable with mixed axion/axino dark matter.
Since they predict $m_{\tg}\alt 500$ GeV (to allow for Yukawa coupling
unification), they will be ruled in or out by LHC in 2011.
\ei

\bigskip

{\it Acknowledgements}:
I thank V. Barger, A. Box, S. Kraml, A. Lessa, A. Mustafayev, E. K. Park, 
S. Sekmen, H. Summy, and X. Tata 
for fruitful collaborations, and I thank the organizers
of IDM2010 for creating an excellent conference!

\end{document}